\documentclass[11pt]{article}
\usepackage{latexsym}
\usepackage{amssymb}
\usepackage{epsf}
\usepackage{amsmath}
\usepackage[hypertex]{hyperref}
\usepackage{graphicx}
\usepackage{color}

\def\a{\alpha}

\def\d{\delta}

\def\k{\kappa}

\def\m{\mu}
\def\n{\nu}

\def\s{\sigma}
\def\t{\tau}

\def\D{\Delta}
\def\G{\Gamma}

\def\L{\Lambda}

\newcommand{\gsim}{\gtrsim}

\begin{document}
\pagestyle{empty}

\begin{flushright}
KEK-TH-1731
\end{flushright}

\vspace{3cm}

\begin{center}

{\bf\LARGE Neutrinoful Universe} 
\\

\vspace*{1.5cm}
{\large 
Tetsutaro Higaki$^1$, Ryuichiro Kitano$^{1,2}$, and Ryosuke Sato$^1$
} \\
\vspace*{0.5cm}

{\it
$^1$Institute of Particle and Nuclear Studies,\\
High Energy Accelerator Research Organization (KEK)\\
Tsukuba 305-0801, Japan\\
$^2$The Graduate University for Advanced Studies (Sokendai)\\
Tsukuba 305-0801, Japan\\
}

\end{center}

\vspace*{1.0cm}

\begin{abstract}
{\normalsize
The Standard Model of particle physics fails to explain the important
 pieces in the standard cosmology, such as inflation, baryogenesis, and
 dark matter of the Universe. We consider the possibility that the
 sector to generate small neutrino masses is responsible for all of
 them; the inflation is driven by the Higgs field to break $B-L$ gauge
 symmetry which provides the Majorana masses to the right-handed
 neutrinos, and the reheating process by the decay of the $B-L$ Higgs
 boson supplies the second lightest right-handed neutrinos whose CP
 violating decays produce $B-L$ asymmetry, {\it \`a la}, leptogenesis. The
 lightest right-handed neutrinos are also produced by the reheating
 process, and remain today as the dark matter of the Universe.
In the minimal model of the inflaton potential, one can set the
 parameter of the potential by the data from CMB observations including
 the BICEP2 and the Planck experiments. In such a scenario, the mass of the dark
 matter particle is predicted to be of the order of PeV. 
We find that the decay of the PeV right-handed neutrinos can explain the
 high-energy neutrino flux observed at the IceCube experiments if the
 lifetime is of the order of $10^{28}$~s.
}
\end{abstract} 

\newpage
\baselineskip=18pt
\setcounter{page}{2}
\pagestyle{plain}
\baselineskip=18pt
\pagestyle{plain}

\setcounter{footnote}{0}

\section{Introduction}

Various cosmological observations are telling us that the Standard Model
of particle physics needs some extension.
The observation of the cosmic microwave background (CMB) and its
anisotropy strongly supports the inflationary cosmology \cite{Guth:1980zm, Starobinsky:1980te, Sato:1980yn}, which requires
a process to generate the Standard Model particles after the inflation
era. The particle-antiparticle asymmetry should also be generated after
or during the reheating process. Also, the dark matter of the Universe
must also be produced in the course of the cosmological history. The
Standard Model should be extended to accommodate the inflation,
baryogenesis and dark matter of the Universe.

One of the clues towards the mysteries of the Universe may be the finite
neutrino masses, which are another evidence to go beyond the Standard
Model.
Once three kinds of right-handed neutrinos are introduced in the same way as other fermions,
the global $U(1)_{B-L}$ symmetry becomes non-anomalous,
and thus can be promoted to a Higgsed gauge symmetry.
It seems that all the ingredients to accommodate the
realistic cosmology are present in this $U(1)_{B-L}$ extended Standard Model.

The inflation can be driven by the Higgs field to break $U(1)_{B-L}$ gauge
symmetry~\cite{Okada:2013vxa,Okada:2014lxa} by assuming an appropriate form of the potential
based on the idea of the chaotic inflation \cite{Linde:1983gd}.
After the inflation, the $B-L$ Higgs field oscillates about the minimum
of the potential where $U(1)_{B-L}$ is broken.
The spontaneous breaking of the $B-L$ symmetry can give 
Majorana masses to the right-handed neutrinos through the Yukawa
coupling, explaining the smallness of the neutrino masses by the seesaw
mechanism \cite{Minkowski:1977sc}.
The very same coupling allows the decay of the inflaton oscillation into
the right-handed neutrinos to reheat the Universe. The subsequent decay
of the right-handed neutrinos can provide the baryon asymmetry of the
Universe by the leptogenesis mechanism~\cite{Fukugita:1986hr}. The
lightest right-handed neutrino should also be produced by the inflaton
decay. If it is long-lived, this non-thermal component is a good
candidate of the dark matter of the Universe.

There have been other minimalistic approaches to the connection between
particle physics and cosmology. An realistic model with the minimal
particle content has been constructed in Ref.~\cite{Davoudiasl:2004be},
where the inflaton and the dark matter particle are both introduced as
new scalar fields.
The possibility of the inflaton as the Higgs-like field, thus playing
important roles both in particle physics and cosmology, has been
considered in Refs.~\cite{Shafi:1983bd,Nakayama:2010sk,Hamada:2014iga}.
%
The dark matter of the Universe as the right-handed neutrino has also
been considered in
Refs.~\cite{Peebles:1982ib,Dodelson:1993je,Asaka:2005an,Kusenko:2010ik}
where the mass range of ${\cal O}$(keV) are assumed.

In this paper, we consider the $U(1)_{B-L}$ extended Standard Model
which covers the shortages in the Standard Model including the small
neutrino masses as well as cosmological observations.
We find that this minimalistic scenario is consistent with various
observations such as tensor-to-scalar ratio, spectral index of the CMB
fluctuations, the neutrino masses, baryon asymmetry of the Universe, and
the energy density of the dark matter.
%
We find, in the case where the reheating process is dominated by the
decay of inflaton into the second lightest right-handed neutrinos, the
mass of the dark matter particle is predicted to be of the order of PeV.

%
Since there is no reason to assume that the dark matter particle, the
lightest right-handed neutrino, to be absolutely stable, we expect the
decay of the dark matter to happen 
occasionally somewhere in the Universe.
Through the dimension-four Yukawa interactions, the
main decay mode would be into a lepton and a $W$ boson, or a neutrino
and a $Z/h$ boson.
We demonstrate that the PeV neutrino events found at the IceCube
experiment~\cite{Aartsen:2013bka, Aartsen:2013jdh} can be explained by
the decaying right-handed neutrinos if the lifetime is of the order of $10^{28}$~s\footnote{
See, {\it e.g.}, Refs.~\cite{Feldstein:2013kka,Esmaili:2013gha} for studies on PeV decaying dark matter.
}.

In the following sections, based on the above scenario with $U(1)_{B-L}$
extended Standard Model, we discuss the neutrino flavor structure, an
inflation model with the $B-L$ Higgs, the non-thermal leptogenesis, the
dark matter abundance produced by the decay of the inflaton, and the
signals of decaying right-handed neutrinos at the IceCube experiment.

\section{Model}
\label{sec:model}

We extend the gauge group of the Standard Model into,
\begin{align}
 &SU(3)_c \times SU(2)_L \times U(1)_Y \times U(1)_{B-L}, \nonumber
\end{align}
and introduce the right-handed neutrinos, $N_i$ $(i=1,2,3)$, and the
$U(1)_{B-L}$ Higgs field $\phi_{B-L}$ which is neutral under the
Standard Model gauge group and has charge $-2$ under $U(1)_{B-L}$. The
$U(1)_{B-L}$ symmetry is gauged, and thus the spontaneous breaking of
$U(1)_{B-L}$ would not leave the massless Nambu-Goldstone boson.
%
The following interaction terms are added to the Standard Model:
\begin{align}
 {\cal L}_{\rm int} =&
 y_\nu^{ij} \bar N_i P_L (\ell_j \cdot \tilde H)  + {\rm h.c.}
\nonumber \\
&
+ {\lambda_i \over 2} \phi_{B-L} \bar N_i P_L N_i^c + {\rm h.c.},
\label{eq:Lint}
\end{align}
where $\ell_i$ and $N_i$ are four-component Weyl fermions, {\it i.e.,}
$P_L \ell_i = \ell_i$, $P_R N_i = N_i$. The coupling
constant $\lambda_i$ can be taken to be real and positive without loss
of generality, and the components of $y_\nu^{ij}$, in general, are complex
valued. The potential terms for $\phi_{B-L}$ field can be written as,
\begin{align}
 V (\phi)& = 
 {\kappa \over 4} (| \phi_{\rm B-L} |^2 - v_{B-L}^2)^2
= {\kappa v_{B-L}^4 \over 4}
\left({| \phi_{\rm B-L} |^2 \over v_{B-L}^2} - 1 \right)^2 .
\label{eq:potential}
\end{align}
There can also be an interaction term such as,
\begin{align}
 {\cal L}_{\phi H}& = \kappa' |\phi_{B-L}|^2 |H|^2 .
\label{eq:higgs}
\end{align}
For $v_{B-L} \gsim 5M_{\rm Pl}$ which we assume later, the coupling constant $\kappa'$ is extremely small if we demand this
term would not contribute significantly to the Higgs potential.

The spontaneous breaking of $U(1)_{B-L}$ by $\langle \phi_{B-L} \rangle = v_{B-L}$ generates masses of $N_i$:
\begin{align}
 M_i& = \lambda_i v_{B-L}.
\end{align}
The neutrino masses are, in turn, generated by the seesaw mechanism:
\begin{align}
 m_{\nu}^{ij}& =  y_\nu^{ki} M_k^{-1} y_\nu^{kj} \langle H \rangle^2.
\end{align}

\begin{figure}[t]
\begin{center}
 \includegraphics[width=8cm]{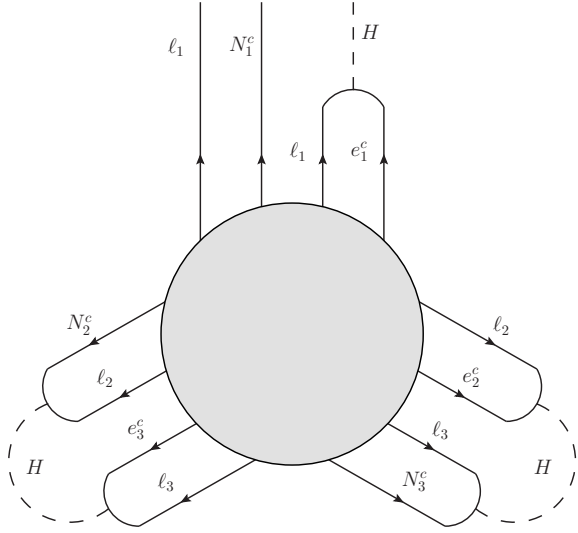}
\end{center}
\caption{One of the diagrams generating $y_\nu^{1 k}$ with the interaction in Eq.~(\ref{determ-int}).} 
\label{fig:nonpert}
\end{figure}

We assume that the lightest right-handed neutrino, $N_1$, to be
long-lived, and it serves as the dark matter of the Universe. That
means,
\begin{align}
 |y_{\nu}^{1i}|&  \ll 1.
\end{align}
As we will see in Sec.~\ref{sec:icecube},
in the scenario where the PeV neutrino events at the IceCube experiment
to be explained by the decay of $N_1$, the lifetime of $N_1$ has to be around $10^{28}$~s.
This lifetime corresponds to $y_\nu^{1i} \sim 10^{-29}$.
In fact, this model has various unexplained small numbers such as the
Higgs mass parameter, the $\theta$ parameter in QCD, the cosmological constant, $\kappa'$, $\kappa$
as well as $y_\nu^{1i}$.
Although we do not look for particular reasons for such small numbers
here, a very small $y_\nu^{1i}$ is somewhat special since it can be
protected by a $Z_2$ symmetry, $N_1 \leftrightarrow - N_1$. If such a
symmetry is only violated by some non-perturbative effects of gauge or
gravity interactions at high scales, the size may be understood as a
natural value~\footnote{The quantum theory of gravity may give natural
ground for such considerations
\cite{Banks:2010zn,BerasaluceGonzalez:2011wy}.}.  In such a scenario, it
is likely that the non-perturbative effects respect the flavor symmetry,
and thus the effective operator to break the $Z_2$ symmetry, for
example, takes the form of
\begin{align}
{\cal L}_{NP} = 
\frac{1}{\L^{14}} (\ell_1 \cdot \ell_2) (\ell_2 \cdot \ell_3) (\ell_3 \cdot \ell_1) 
e^c_1 e^c_2 e^c_3 
N^c_1 N^c_2 N^c_3 + {\rm h.c.}
&
\label{determ-int}
\end{align}
Here, $\Lambda$ is expected to be the scale which characterizes the
non-perturbative effects such as, $\mu e^{-8\pi^2/g^2(\mu)}$, in the
case of a gauge theory. This is analogous to the interaction considered
in QCD \cite{'tHooft:1976fv}.
Together with the Yukawa interactions of the charged lepton sector
$y_e^{ij}$ in the Standard Model and $y_\nu^{\alpha i}~(\alpha = 2,3)$
in Eq.~\eqref{eq:Lint}, $y_\nu^{1i}$ is generated as in the diagram in
Fig.~\ref{fig:nonpert}:
\begin{align}
y_\nu^{1 k} \propto (\det y_e) \epsilon^{ijk} y_\nu^{2 i} y_\nu^{3 j}. &
\end{align}
One can also consider interactions such as ${\cal L}_{NP} = 
(q_1 \cdot \ell_2) (q_2 \cdot \ell_3) (q_3 \cdot \ell_1) d^c_1 d^c_2
d^c_3 N^c_1 N^c_2 N^c_3 / \Lambda^{14} $.  From this operator, we obtain $y_\nu^{1 k}
\propto (\det y_d) \epsilon^{ijk} y_\nu^{2 i} y_\nu^{3 j}$. 
In any case, the flavor symmetry implies an interesting proportionality:
\begin{align}
 y_\nu^{1k}& \propto \epsilon^{ijk} y_\nu^{2 i} y_{\nu}^{3 j}.
\label{eq:prop}
\end{align}
%
%
We will see in Sec.~\ref{sec:icecube} that if this type of
contribution is dominated, the branching ratio of the $N_1$ decay is
directly related to the neutrino mixing parameters. By introducing a
small parameter $c$, Eq.~\eqref{eq:prop} is explicitly written as
\begin{align}
y^{1e}_\n = c (y^{2\m}_\n y^{3\t}_\n - y^{3\m}_\n y^{2\t}_\n), \quad
y^{1\m}_\n = c (y^{2\t}_\n y^{3e}_\n - y^{3\t}_\n y^{2e}_\n), \quad
y^{1\t}_\n = c (y^{2e}_\n y^{3\m}_\n - y^{3e}_\n y^{2\m}_\n). \label{eq:DMneutrinoyukawa}
\end{align}

Because of tiny $y_\n^{1\ell}$'s, $N_1$ provides very little
contribution to the neutrino masses.  In this case, the neutrino sector
is essentially that of the model with only two right-handed
neutrinos~\cite{Frampton:2002qc, Harigaya:2012bw}.
Here, we define the following Yukawa matrix $\tilde y$ and mass matrix
$\tilde M$:
\begin{align}
{\tilde y} = \left(\begin{array}{ccc}
y^{2e}_\n & y^{2\m}_\n & y^{2\t}_\n \\
y^{3e}_\n & y^{3\m}_\n & y^{3\t}_\n
\end{array}\right),\qquad
{\tilde M} = \left(\begin{array}{cc}
M_2 & 0 \\
0 & M_3 
\end{array}\right).
\end{align}
Neutrino masses are given by,
\begin{align}
m_\n \equiv {\rm diag}(m_1, m_2, m_3) = (U_{\rm PMNS}^T {\tilde y}^T
 {\tilde M}^{-1} {\tilde y} U_{\rm MNS}) \langle H\rangle^2, \label{eq:neutrinomass}
\end{align}
where $U_{\rm MNS}$ is the Pontecorvo-Maki-Nakagawa-Sakata (PMNS) matrix \cite{Pontecorvo:1957qd,MNS}:
\begin{align}
U_{\rm PMNS} =
\left(\begin{array}{ccc}
U_{ e 1} & U_{ e 2} & U_{ e 3}\\
U_{\m 1} & U_{\m 2} & U_{\m 3}\\
U_{\t 1} & U_{\t 2} & U_{\t 3}
\end{array}\right) 
=&
\left(\begin{array}{ccc}
 c_{12}c_{13}                           &  s_{12}c_{13}                           & s_{13}e^{-i\d} \\
-s_{12}c_{23}-c_{12}s_{23}s_{13}e^{i\d} &  c_{12}c_{23}-s_{12}s_{23}s_{13}e^{i\d} & s_{23}c_{13}   \\
 s_{12}s_{23}-c_{12}c_{23}s_{13}e^{i\d} & -c_{12}s_{23}-s_{12}c_{23}s_{13}e^{i\d} & c_{23}c_{13}
\end{array}\right) \nonumber\\
& \times {\rm diag}(1,e^{i\a/2},1).\label{eq:mns}
\end{align}
Eq.~(\ref{eq:neutrinomass}) tells us that the lightest neutrino is
massless (up to $O((y_\nu^{1i})^2)$ contributions) because the rank of
$\tilde y$ and $\tilde M$ is two. There is only one Majorana phase in
Eq.~(\ref{eq:mns}) in this effectively two-generation model.  We can
parametrize $\tilde y$ which satisfies Eq.~(\ref{eq:neutrinomass}) by
using a $3\times 2$ complex matrix $R$ \cite{Casas:2001sr, Ibarra:2003up}:
\begin{align}
\tilde y = \frac{1}{\langle H \rangle} {\tilde M}^{1/2} R m_\n^{1/2}
 U_{\rm PMNS}^\dagger, \label{eq:neutrinoyukawa}
\end{align}
where $R$ can be expressed in terms of a complex parameter $z$,
\begin{align}
R=\left(\begin{array}{ccc}
0 & \cos z & -\sin z\\
0 & \sin z & \cos z
\end{array}\right), \label{eq:R_normal}
\end{align}
for normal hierarchy, and,
\begin{align}
R=\left(\begin{array}{ccc}
\cos z & -\sin z & 0\\
\sin z & \cos z & 0
\end{array}\right), \label{eq:R_inverted}
\end{align}
for inverted hierarchy.

By using the above parametrization and Eqs.~(\ref{eq:DMneutrinoyukawa}, \ref{eq:neutrinoyukawa}),
we can determine the structure of the Yukawa coupling $y_\n$.
For normal hierarchy, we obtain,
\begin{align}
y^{1\ell}_\n &= c \frac{\sqrt{M_2 M_3 m_2 m_3}}{\langle H\rangle^2} \det U_{\rm PMNS}^* \times U_{\ell 1},\\
y^{2\ell}_\n &= \frac{\sqrt{M_2}}{\langle H\rangle}\left( \sqrt{m_2} U_{\ell 2}^* \cos z - \sqrt{m_3} U_{\ell 3}^* \sin z \right),\\
y^{3\ell}_\n &= \frac{\sqrt{M_3}}{\langle H\rangle}\left( \sqrt{m_2} U_{\ell 2}^* \sin z + \sqrt{m_3} U_{\ell 3}^* \cos z \right).
\end{align}
For inverted hierarchy,
\begin{align}
y^{1\ell}_\n &= c \frac{\sqrt{M_2 M_3 m_1 m_2}}{\langle H\rangle^2} \det U_{\rm PMNS}^* \times U_{\ell 3},\\
y^{2\ell}_\n &= \frac{\sqrt{M_2}}{\langle H\rangle}\left( \sqrt{m_1} U_{\ell 1}^* \cos z - \sqrt{m_2} U_{\ell 2}^* \sin z \right),\\
y^{3\ell}_\n &= \frac{\sqrt{M_3}}{\langle H\rangle}\left( \sqrt{m_1} U_{\ell 1}^* \sin z + \sqrt{m_2} U_{\ell 2}^* \cos z \right).
\end{align}
Here, we used the unitarity of $U_{\rm PMNS}$ for the calculation of $y_\n^{1\ell}$.
These structures are important for the discussion of the flavor of the
decay products of $N_1$, We will discuss their effects on the energy
spectrum of the neutrino flux from the decay of $N_1$ in
Sec.~\ref{sec:icecube}.

\section{Inflation with the $B-L$ Higgs field}
In this section, we consider an inflation model with the $B-L$ Higgs field.
The potential for $\phi_{B-L}$ in Eq.~\eqref{eq:potential} can drive
inflation of the Universe.
By defining $\phi = \sqrt{2} |\phi_{B-L}|$, the potential is recast in
the form of,
\begin{align}
 V (\phi)& = 
\Lambda^4 \left(
{\phi^2 \over \mu^2} - 1 \right)^2 ,
\end{align}
where $\mu^2 = 2 v_{B-L}^2$ and $\Lambda^4 = \kappa v_{B-L}^4 / 4$, and
we define $\mu > 0$. The phase direction can be gauged away.

The inflaton field can slow roll when $\mu \gg M_{\rm Pl}$, either from
the $|\phi| > \mu$ or $|\phi| < \mu$ region towards the minimum at $\phi
= \mu$.
%
%
In both cases, the slow-roll parameters at $\phi = \phi_0$ are given by \cite{Liddle:2000cg},
\begin{align}
 \epsilon& = {M_{\rm Pl}^2 \over 2} \left(
{V' \over V}
\right)^2
= {M_{\rm Pl}^2 \over 2 \mu^2} \left(
{4\phi_0 \over \mu} \over {{\phi_0^2 \over \mu^2} - 1}
\right)^2, \quad
 \eta  =
M_{\rm Pl}^2 {V'' \over V}
= {4M_{\rm Pl}^2 \over \mu^2}
{
{3 \phi_0^2 \over \mu^2} - 1
\over 
\left({\phi_0^2 \over \mu^2} - 1\right)^2 
}.
\end{align}
The field value $\phi_0$ at the pivot scale $k_0 = 0.002$~Mpc$^{-1}$ is
expressed in terms of the number of $e$-folds $N$:
\begin{align}
 N & \simeq {1 \over M_{\rm Pl}^2} \int_{\phi_{\rm end}}^{\phi_0}
{V \over V'} d\phi
=
{\mu \over M_{\rm Pl}^2} \int_{\phi_{\rm end}}^{\phi_0}
{
{{\phi^2 \over \mu^2} - 1} \over {4\phi \over \mu}
}
 d\phi,
\end{align}
where the field value at the end of the inflation, $\phi_{\rm end}$, is
obtained from,
\begin{align}
 1 &\simeq M_{\rm Pl}^2 \left(
{V' \over V}
\right)^2 \Bigg|_{\phi = \phi_{\rm end}}
= {M_{\rm Pl}^2 \over  \mu^2} \left(
{4\phi_{\rm end} \over \mu} \over {{\phi_{\rm end}^2 \over \mu^2} - 1}
\right)^2.
\end{align}
The tensor-to-scalar ratio, $r$, and the spectral index, $n_s$, is
expressed in terms of the slow-roll parameters as,
\begin{align}
 r = 16 \epsilon,\qquad
 n_s = 1 - 6 \epsilon + 2 \eta.
\end{align}
The Planck normalization sets the overall scale \cite{Ade:2013uln},
\begin{align}
 \left({V \over \epsilon}\right)^{1/4} \Bigg|_{\phi_0}& = 6.4 \times 10^{16}~{\rm GeV},
\end{align}
and the observed spectral index is given by,
\begin{align}
n_s & = 0.9603 \pm 0.0073 .
\end{align}
The results from the BICEP2 experiment prefer,
\begin{align}
r  = 0.20^{+0.07}_{-0.05} , \qquad
V^{1/4}  = 2.0 \times 10^{16} {\rm GeV} \cdot \bigg(\frac{r}{0.16}\bigg)^{1/4},
\end{align}
when one combines the data from the Planck experiment.\footnote{ We will
not consider the tension between the data from the Planck satellite ($r
< 0.11$) \cite{Ade:2013uln} and that from the BICEP2 experiment ($r \sim
0.2$).  The tension can be relaxed if one considers a running spectral
index, an extra relativistic component, non-zero neutrino
mass~\cite{Ade:2014xna,Giusarma:2014zza}, an anti-correlation between
tensor and scalar modes~\cite{Contaldi:2014zua} or between tensor and
isocurvature modes~\cite{Kawasaki:2014lqa}.  See also
\cite{Miranda:2014wga,Freivogel:2014hca} for other solutions.  }  Here,
the preferred range of $r$ will be modified to $r =
0.16^{+0.06}_{-0.05}$ after subtracting the best available estimate for
foreground dust \cite{Ade:2014xna}.

\begin{figure}[t]
\begin{center}
 \includegraphics[width=7cm]{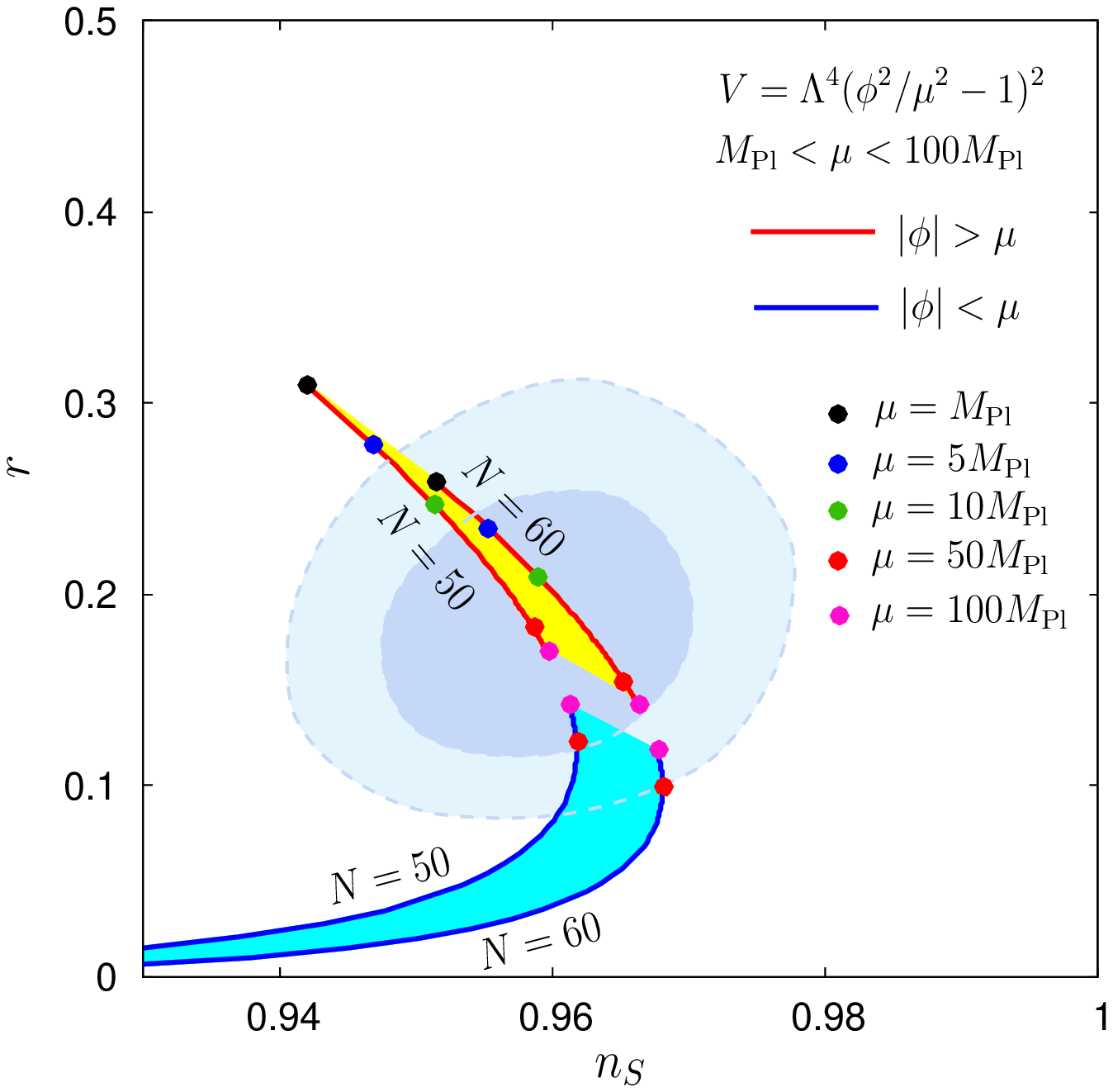}
\hspace{0.5cm}
 \includegraphics[width=7.3cm]{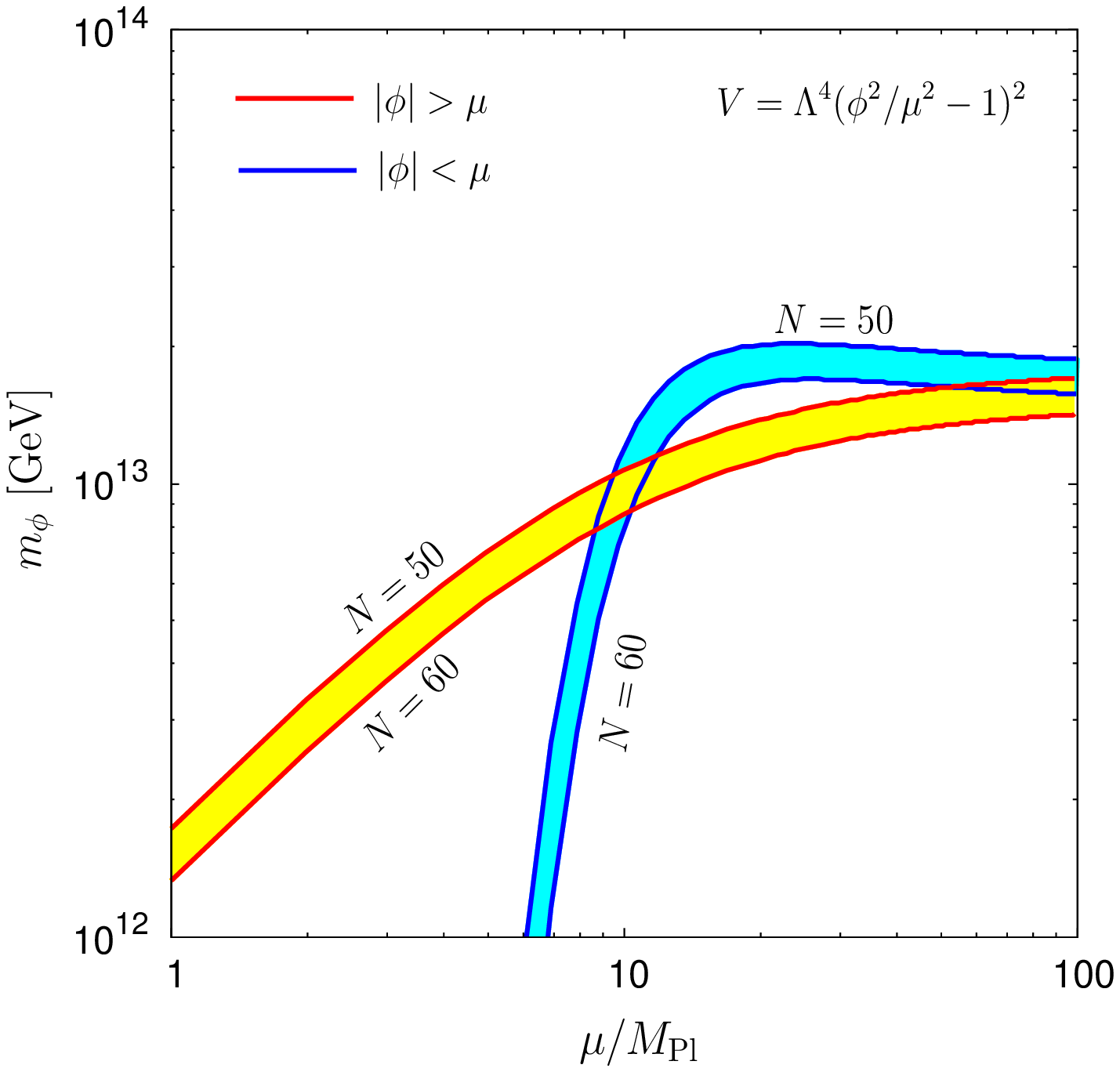}
\end{center}
\caption{Predictions of the inflation model in the $r-n_s$ plane (left)
 and $m_\phi - \mu/M_{\rm Pl}$ plane (right). The
region favored by CMB observations (Planck+WP+highL+BICEP2) \cite{Ade:2014xna} are also
shown in the left figure. The dark blue region corresponds to the region consistent with the BICEP2 at $1\sigma$ level 
whereas the light blue region does to that at 2$\sigma$ level.
The left figure is consistent with the result obtained in Ref.\cite{Okada:2014lxa}.} 
\label{fig:ns-r}
\end{figure}

The predictions for $r$ and $n_s$ is shown in Fig.~\ref{fig:ns-r} with
varying $\mu$. The region favored by the CMB observations are also
shown. We see that for $N=60$, $|\phi|>\mu$ and $\mu \gtrsim 5 M_{\rm
Pl}$ is favored.  The inflaton mass, $m_\phi = 2 \sqrt{2} \Lambda^2 /
\mu$, as a function of $\mu/M_{\rm Pl}$ is also shown in
Fig.~\ref{fig:ns-r}.  For $|\phi| > \mu$ and $\mu \gtrsim 5 M_{\rm Pl}$,
we find,
\begin{align}
 m_\phi& \sim 10^{13}~{\rm GeV}.
\end{align}
This value corresponds to a very small value of $\k$ such as $\k \sim 10^{-12}$ for $\m \sim 5 M_{\rm Pl}$.
In the following discussion, we fix the inflaton mass at this value, and
will see that the correct amount of the baryon asymmetry and the dark
matter can be obtained after the decay of the inflaton fields.

\section{Reheating by the inflaton decay}

%
After the inflation, the decay of $\phi$ can produce the Standard Model
particles. The dominant decay mode can either be into two right-handed
neutrinos via the interaction term in Eq.~\eqref{eq:Lint} or two Higgs
fields (including the Goldstone modes) via the term in
Eq.~\eqref{eq:higgs}.

In the case where the $\phi \to N_i N_i$ mode is dominated and for
$\lambda_1 \ll \lambda_2$ and $M_3 > m_\phi$ which are justified later,
the decay width is given by,
%
%
\begin{align}
 \Gamma_\phi & = {1 \over 2} 
{m_\phi \over 16 \pi}
{M_2^2 \over v_{B-L}^2}
\left(
1 - {4 M_{2}^2 \over m_\phi^2}
\right)^{3/2}.
\end{align}
By equating $\Gamma_\phi$ with the Hubble parameter $H(T_R)$ at the
reheating temperature, $T_R$, we obtain,
\begin{align}
 T_R& \simeq 2 \times 10^{7}~{\rm GeV}
\left(
{M_2 \over 10^{12}~{\rm GeV}}
\right)
\left(
{m_\phi \over 10^{13}~{\rm GeV}}
\right)^{1/2}
\left(
{v_{B-L} \over 5 M_{\rm Pl}}
\right)^{-1}
\left(
1 - {4 M_2^2 \over m_\phi^2}
\right)^{3/4}.
\end{align}
Here we used $T_R = (90/\pi^2 g_*(T_R))^{1/4}\sqrt{\Gamma_{\phi}M_{\rm
Pl}}$ and $g_*(T_R) = 106.75$, where $g_*(T_R)$ is the relativistic
degrees of freedom in plasma at the temperature $T_R$.  If the Higgs
mode $\phi \to hh, WW, ZZ$ is the dominant decay channel through
Eq.~(\ref{eq:higgs}), the reheating temperature can be arbitrarily
higher than the above estimate. If $T_R$ is higher than $m_\phi$, the
perturbative analysis of the reheating process becomes unreliable
\cite{Mukaida:2012qn}. Therefore, we restrict ourselves to the region of
$T_R < m_\phi \sim 10^{13}$~GeV.


\subsection{Leptogenesis}

For the case where $\phi \to N_2 N_2$ is the dominant decay mode,
the decay of $N_2$ can generate $B-L$ asymmetry by leptogenesis.
The baryon-to-entropy ratio obtained from the non-thermal leptogenesis is \cite{Asaka:1999yd},
\begin{align}
 {n_B \over s}& = 
-{28 \over 79}
\cdot {3 \over 2} 
\cdot \epsilon 
\cdot {T_R \over m_\phi},
\end{align}
where $(3/2) T_R / m_\phi$ is the number density $n_{N_2} \simeq n_\phi /2$ divided by the
entropy density produced by the decay of $\phi$.
The $\epsilon$ factor is the magnitude of the CP violation \cite{Covi:1996wh}:
\begin{align}
 \epsilon& \simeq -{3 \over 16 \pi}{
{\rm Im} (y_\nu y_\nu^\dagger)_{23}^2
\over
(y_\nu y_\nu^\dagger)_{22}
} {M_2 \over M_3},
\end{align}
for $M_2 \ll M_3$.
It is bounded by \cite{ Harigaya:2012bw, Davidson:2002qv},
\begin{align}
| \epsilon |& \lesssim
\left\{\begin{array}{ll}
 \displaystyle{3 \over 16 \pi}{M_2 \over \langle H \rangle^2} (m_3-m_2)
\sim  8 \times 10^{-5} \left(
{{M_2} \over 10^{12}~{\rm GeV}
}
\right), & \quad({\rm Normal})\\
 \displaystyle{3 \over 16 \pi}{M_2 \over \langle H \rangle^2} (m_2-m_1)
\sim  2 \times 10^{-6} \left(
{{M_2} \over 10^{12}~{\rm GeV}
}
\right). & \quad({\rm Inverted})
\end{array}\right.
\end{align}
Here, we take $\D m_{\odot}^2 = (0.0086~{\rm eV})^2$ and $\D m_A^2 = (0.048~{\rm eV})^2$ \cite{Beringer:1900zz}.
 Therefore,
\begin{align}
 {n_B \over s} \Big|_{\rm max}& \simeq 
\left(
{M_2 \over 10^{12}~{\rm GeV}}
\right)^2
\left({m_\phi \over 10^{13}~{\rm GeV}}\right)^{-1/2}
\left({v_{B-L} \over 5 M_{\rm Pl}}\right)^{-1}
\times 
\left\{\begin{array}{ll}
1 \times 10^{-10} & ({\rm Normal})\\
2 \times 10^{-12} & ({\rm Inverted})
\end{array}\right..
\end{align}
For normal hierarchy, compared with the observed baryon-to-entropy
ratio, $n_B / s \simeq 10^{-10}$ \cite{Ade:2013zuv}, we need $M_2
\gtrsim 10^{12}$~GeV.  On the other hand, for inverted hierarchy, we
need $M_2 \gtrsim 10^{13}$~GeV which is on the border of the constraint:
$m_\phi > 2 M_2$.  In any case, these result justify $M_3 > m_\phi$
which we assumed before.

If the Higgs mode is important, the branching ratio into $M_2$ is
suppressed, and thus non-thermal leptogenesis becomes difficult. With
fixed $m_\phi$ from the CMB observations, there is no freedom to make
$M_2$ larger since the decay into $N_2$ becomes kinematically forbidden.
Instead, if the reheating temperature is high enough, it is possible to
produce $N_2$ thermally.  The thermal leptogenesis is possible for
$10^{9}~{\rm GeV} \lesssim M_2 \lesssim T_R$
\cite{Davidson:2002qv,Buchmuller:2002rq}.

%

\subsection{Dark matter abundance}
The inflaton also decays into two $N_1$'s. The assumption that $N_1$ is
long-lived makes it possible to identify this component to be the dark
matter of the Universe.

\begin{figure}[t]
\begin{center}
 \includegraphics[width=10cm]{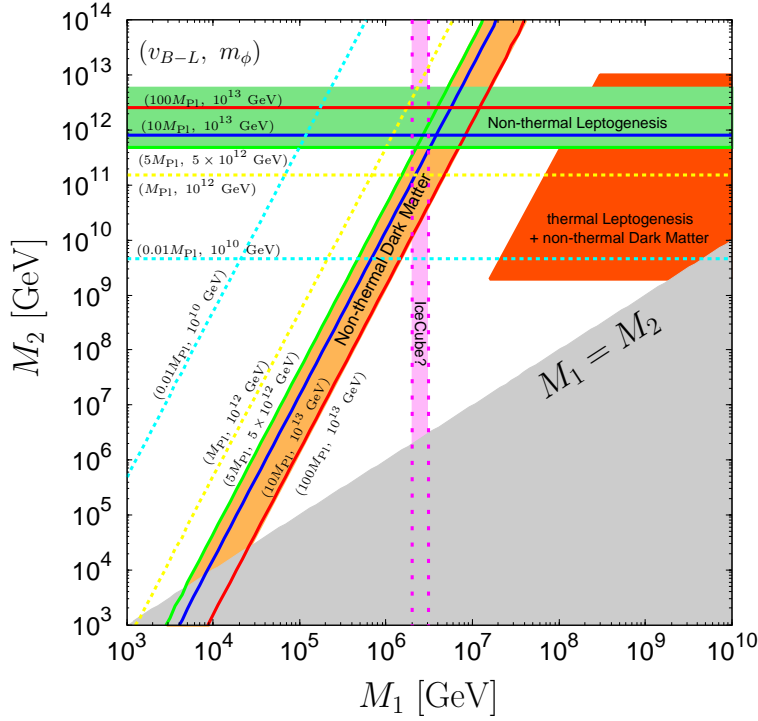}
\end{center}
\caption{Consistent regions with neutrino masses and cosmological
observations.  The two shaded regions (green and light orange) are
consistent with the BICEP2 at $1\sigma$ level respectively, and imply
that the non-thermal leptogenesis works (green) and the dark matter is
explained via the inflaton decay (light orange).  Here, we assume normal
hierarchy.  We also show the mass range of $N_1$ favored by the IceCube
experiment (pink shaded region).  In the dark orange region where
thermal leptogenesis is viable, the reheating temperature is treated as
a free parameter satisfying $M_2 \leq T_R \leq m_{\phi} =10^{13}$ GeV
with $5M_{\rm Pl} \leq v_{B-L}$. A high reheating temperature is
 realized by the decay into $hh$, $WW$ and $ZZ$ via the coupling in
Eq.~(\ref{eq:higgs}).  } \label{fig:consistentregion}
\end{figure}

The partial decay width is given by,
\begin{align}
 \Gamma (\phi \to N_1 N_1) & = {1 \over 2} 
{m_\phi \over 16 \pi}
{M_1^2 \over v_{B-L}^2}
\left(
1 - {4 M_{1}^2 \over m_\phi^2}
\right)^{3/2}.
\end{align}
By using the relation $H(T_R) \sim \Gamma_\phi \sim T_R^2/M_{\rm Pl}$, and $n_{N_1} / s \simeq
(3/2) (T_R / m_\phi) {\rm Br}(\phi \to N_1 N_1)$, we find,
\begin{align}
 {\Omega_{N_1}^{\rm NT}} &\simeq
0.2 \left(
{M_1 \over 4~{\rm PeV}}
\right)^3
\left(
{T_R \over 3 \times 10^{7}~{\rm GeV}}
\right)^{-1}
\left(
{v_{B-L} \over 5 M_{\rm Pl}}
\right)^{-2}.
\end{align}
Here, we used $\Omega_{N_1}^{\rm NT} = (M_1 n_{N_1}/s)/(\rho_c/s)_0$, where $(\rho_c/s)_0 \simeq 1.8 \times 10^{-9}$ GeV is
the critical density divided by the entropy density today.
%
%
%
The contribution from the thermal production from the scattering
processes by the $U(1)_{B-L}$ gauge interaction is much smaller such as \cite{Kusenko:2010ik},
\begin{align}
 \Omega_{N_1}^{\rm TH}& \sim 10^{-23} 
\left(
{M_1 \over 4~{\rm PeV}}
\right)
\left(
{T_R \over 5 \times 10^{7}~{\rm GeV}}
\right)^3
\left(
{v_{B-L} \over 5 M_{\rm Pl}}
\right)^{-4}.
\end{align}
This is estimated with the interaction between $N_1$ and the Standard
Model fermions in plasma through the s-channel exchange of the
$U(1)_{B-L}$ gauge boson.  We summarize the allowed regions in
Fig.~\ref{fig:consistentregion}.  We see that one obtains the correct
amount of the baryon asymmetry and the dark matter abundance at $M_1
\sim 1$ PeV and $M_2 \sim 10^{12}$ GeV within the region consistent with
the BICEP2 results at the $1\sigma$ level.
The PeV dark matter opens up an interesting possibility that the high energy neutrinos observed
at the IceCube experiment \cite{Aartsen:2013bka, Aartsen:2013jdh} are explained by the decay of $N_1$,
which will be studied in the next section.
For a heavier $N_1$, we see a region where the thermal leptogenesis
works. There, a high enough reheating temperature is realized by the
inflaton decay into Higgs fields through the coupling in
Eq.~(\ref{eq:higgs}).

\section{PeV neutrinos as a signal of decaying $N_1$} \label{sec:icecube}

In this section, we discuss observational signatures of the right-handed
neutrino dark matter.  As discussed so far, the inflation, the baryon
asymmetry and the correct amount of dark matter can be explained for
$M_1 = {\cal O}(1)$~PeV.
Since there is no reason to expect that $N_1$ is absolutely stable, we
have a chance to see high energy cosmic rays produced from the decay 
of $N_1$.
It is interesting that the PeV is indeed the energy region where an
excess of high energy neutrinos events are observed at the IceCube
experiment.
In this section, we discuss the possibility that neutrino excess which
is observed at IceCube experiment \cite{Aartsen:2013bka,
Aartsen:2013jdh} is explained by the decay products of $N_1$.

\subsection{The branching fractions of $N_1$}
The partial decay widths of $N_1$ at tree level are,
\begin{align}
\G(N_1\to\ell^- W^+) = \G(N_1\to\ell^+ W^-) &= \frac{|y^{1\ell}_\n|^2 M_1}{16\pi} \left( 1 - \frac{m_W^2}{M_1^2}\right)^2 \left( 1 + \frac{2m_W^2}{M_1^2}\right),\\
\G(N_1\to\n_\ell Z) = \G(N_1\to\bar\n_\ell Z) &= \frac{|y^{1\ell}_\n|^2 M_1}{32\pi} \left( 1 - \frac{m_Z^2}{M_1^2}\right)^2 \left( 1 + \frac{2m_Z^2}{M_1^2}\right),\\
\G(N_1\to\n_\ell h) = \G(N_1\to\bar\n_\ell h) &= \frac{|y^{1\ell}_\n|^2 M_1}{32\pi} \left(1-\frac{m_h^2}{M_1^2}\right)^2.
\end{align}
For $M_1 \gg m_W,~m_Z,~m_h$, we can see that $\G(N_1\to\ell^\mp W^\pm) : \G(N_1\to \n Z,\bar\n Z) : \G(N_1\to \n h,\bar\n h) \simeq 2:1:1$
due to the equivalence theorem \cite{Cornwall:1974km}.
The lifetime of $N_1$ for $M_1 \gg m_W,~m_Z,~m_h$ is calculated as,
\begin{align}
\t_{N_1} =  \left( \frac{M_1}{4\pi} \sum_{\ell} |y_{1\ell}|^2 \right)^{-1}
\sim 8 \times 10^{28}~{\rm s} \left( \frac{M_1}{1~{\rm PeV}}\right)^{-1} \left( \sum_{\ell} \left| \displaystyle\frac{y_{1\ell}}{10^{-29}} \right|^2  \right)^{-1}.
\end{align}
The branching fractions for each lepton family ${\rm Br}(\ell) \equiv {\rm Br}(N_1 \to \ell^\mp W^\pm,~\n_\ell Z,~\bar\n_\ell Z,~\n_\ell h,~\bar\n_\ell h)$
 are determined by $y_\n^{1\ell}$'s.
For each neutrino mass hierarchy, by the assumption of Eq.~(\ref{eq:DMneutrinoyukawa}), ${\rm Br}(\ell)$'s are completely determined by the PMNS matrix,
\begin{align}
( {\rm Br}(e),~{\rm Br}(\m),~{\rm Br}(\t) ) &= (|U_{e1}|^2 ,~ |U_{\m 1}|^2 ,~ |U_{\t 1}|^2), & ({\rm Normal})\\
( {\rm Br}(e),~{\rm Br}(\m),~{\rm Br}(\t) ) &= (|U_{e3}|^2 ,~ |U_{\m 3}|^2 ,~ |U_{\t 3}|^2). & ({\rm Inverted})
\end{align}
We take $\sin^2\theta_{12} =  0.31$, $\sin^2\theta_{23} =  0.39$ and $\sin^2\theta_{13} =  0.02$ \cite{Beringer:1900zz},
then, the numerical values of the branching fraction are given by,
\begin{align}
({\rm Br}(e),~{\rm Br}(\m),~{\rm Br}(\t)) &= (0.68 ,~0.24 + 0.02 \cos\d ,~0.08 - 0.02 \cos\d), &({\rm Normal})\\
({\rm Br}(e),~{\rm Br}(\m),~{\rm Br}(\t)) &= (0.02 ,~0.38 ,~0.60). &({\rm Inverted})
\end{align}
The branching fractions for normal hierarchy has small dependence on
CP-violating phase $\d$.  On the other hand, the branching fractions for
inverted hierarchy is completely determined independent of $\d$.

\subsection{Neutrino flux from decay of $N_1$}

\begin{figure}[p]
\centering
\begin{minipage}{0.49\hsize}
\centering\includegraphics[width=\hsize]{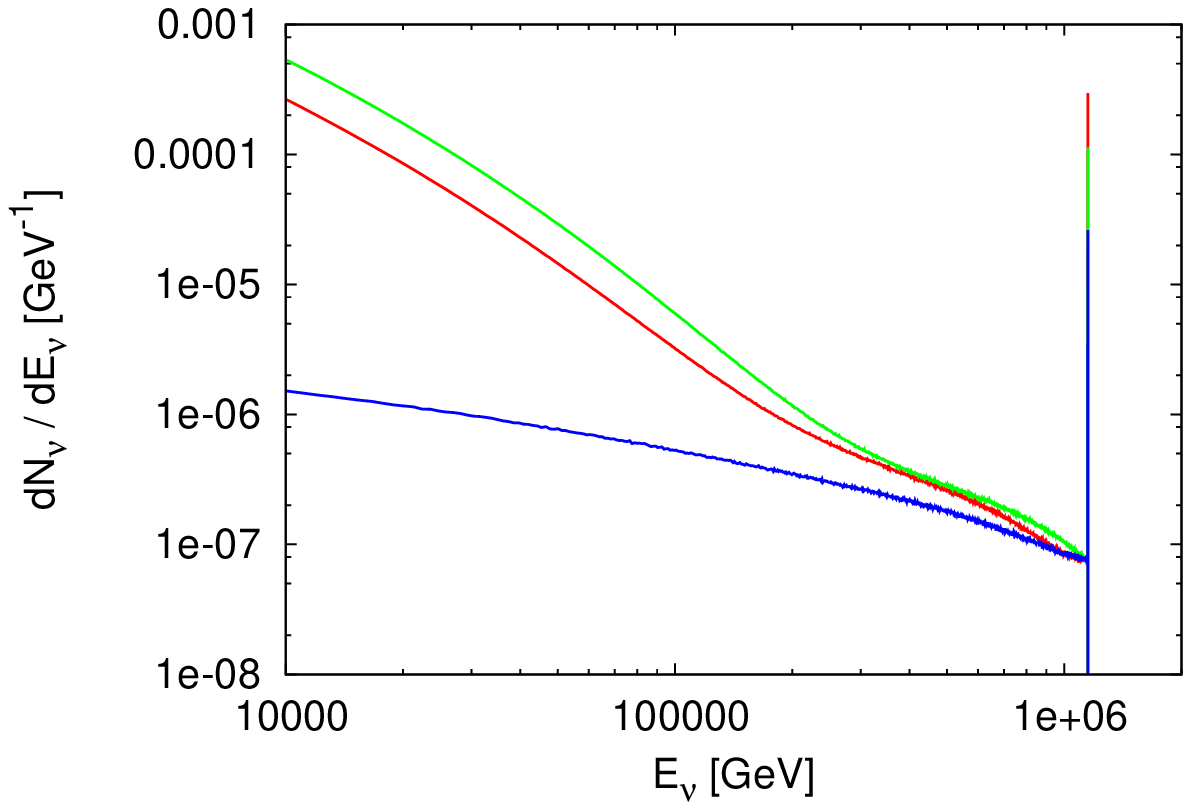}\\
(a) Normal hierarchy ($\d=0$)
\end{minipage}
\begin{minipage}{0.49\hsize}
\centering\includegraphics[width=\hsize]{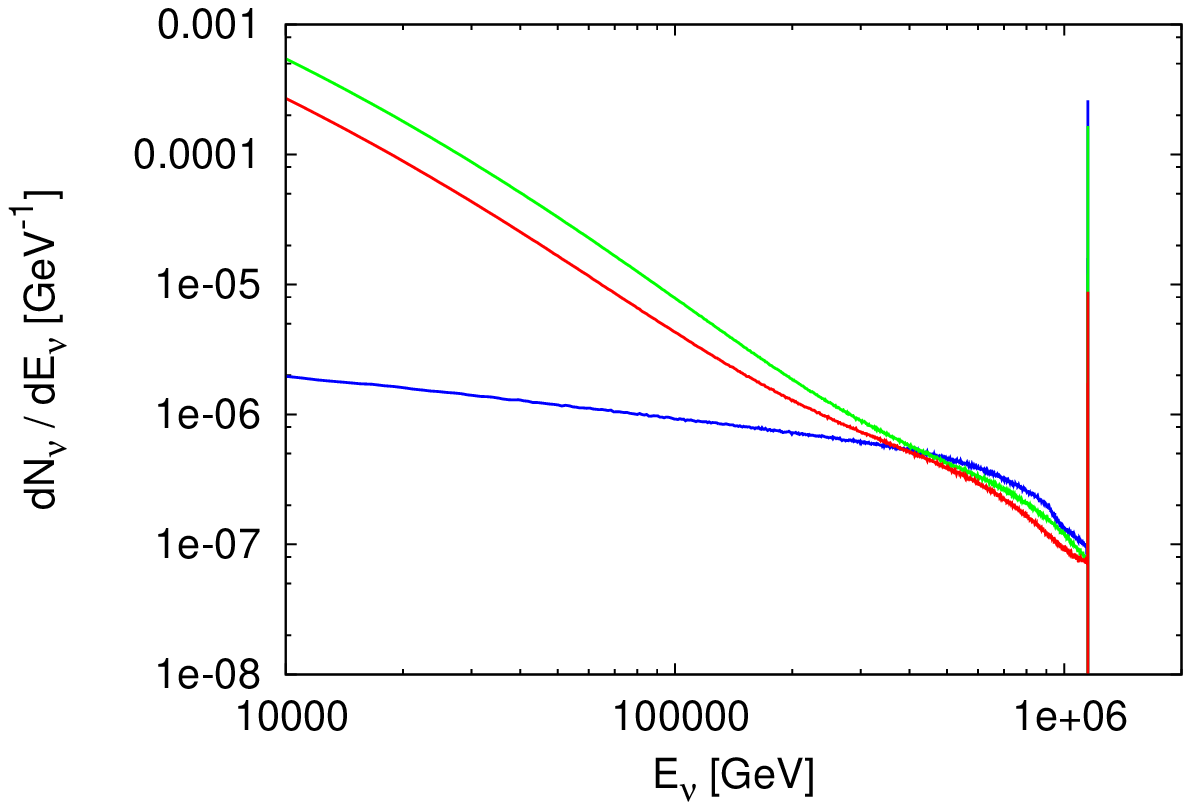}\\
(b) Inverted hierarchy ($\d=0$)
\end{minipage}
\caption{ $dN_\n / dE_\n$ for $M_1=2.3~{\rm PeV}$ when produced by the
decay of $N_1$.  We take normal hierarchy with $\d=0$ in left figure and
inverted hierarchy in right figure.  Red, green and blue lines show the
spectrum of $\n_e + \bar \n_e$, $\n_\m + \bar \n_\m$ and $\n_\t + \bar
\n_\t$, respectively.  }\label{fig:pythia}
\vspace{1cm}~\\
\centering
\begin{minipage}{0.49\hsize}
\centering\includegraphics[width=\hsize]{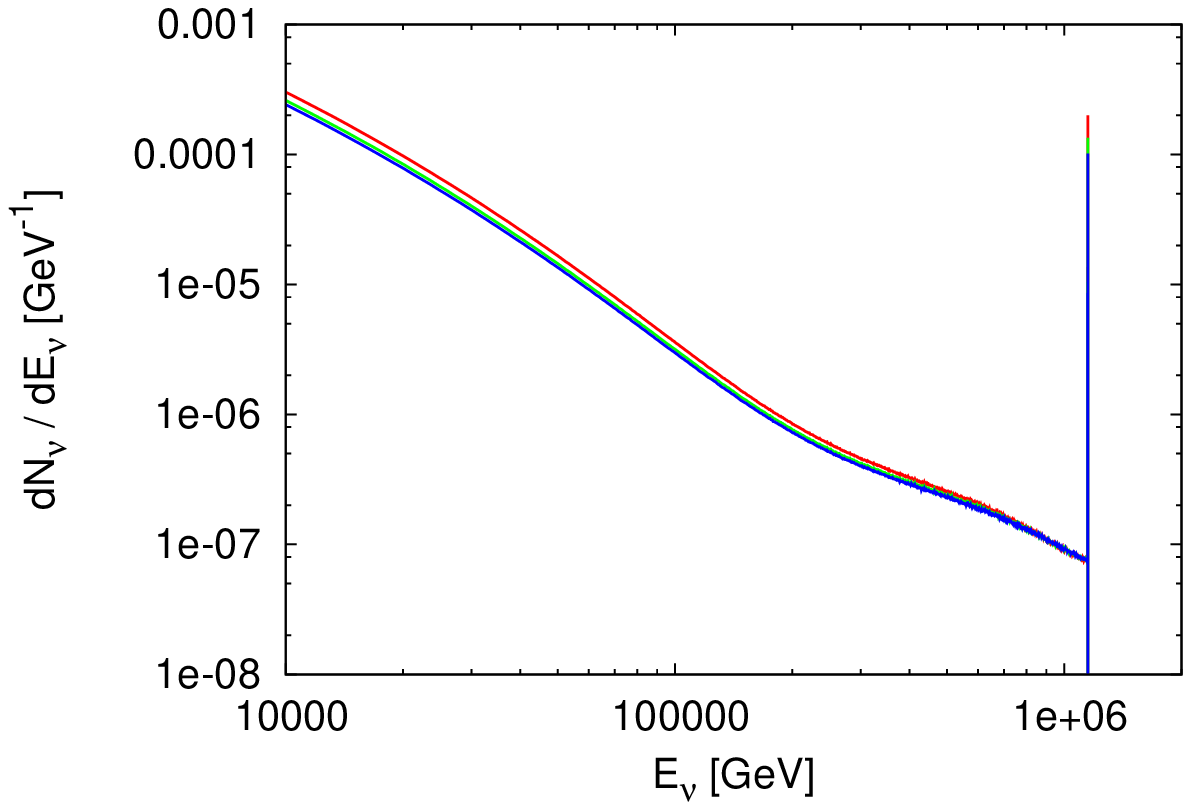}\\
(a) Normal hierarchy ($\d=0$)
\end{minipage}
\begin{minipage}{0.49\hsize}
\centering\includegraphics[width=\hsize]{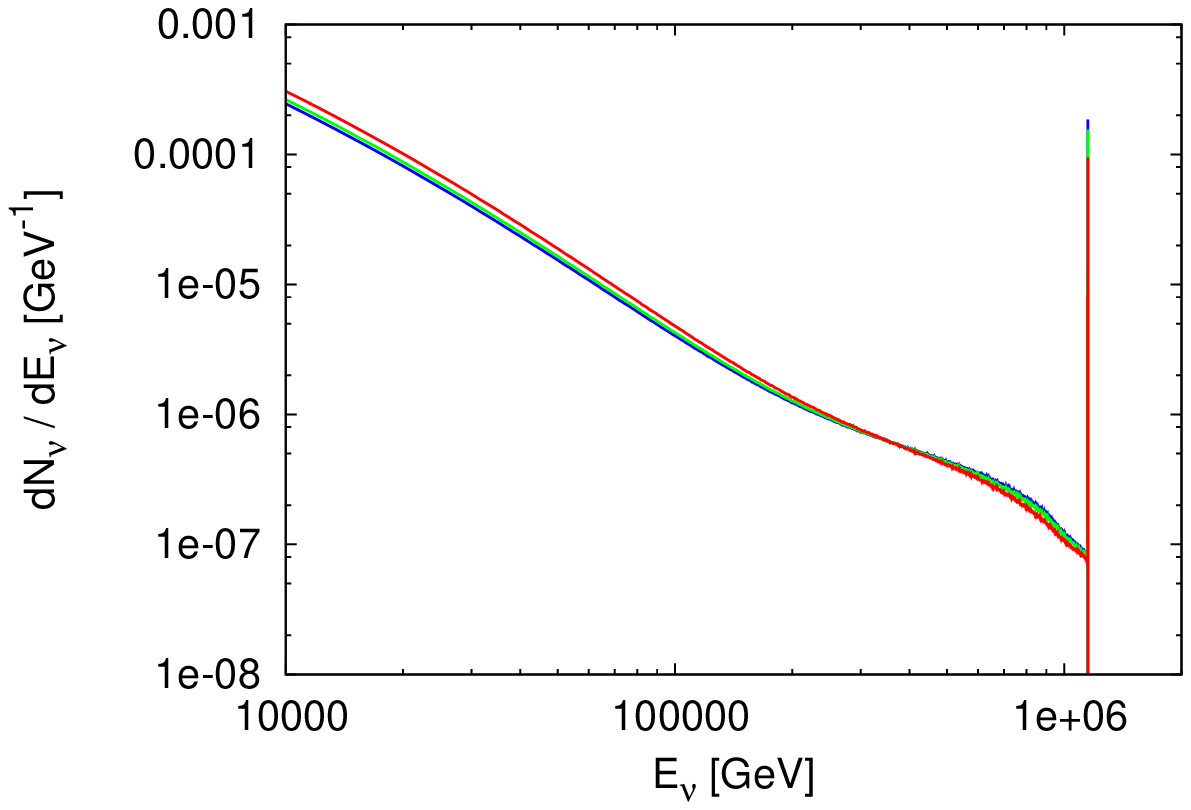}\\
(b) Inverted hierarchy ($\d=0$)
\end{minipage}
\caption{
$dN_\n / dE_\n$ for $M_1=2.3~{\rm PeV}$ which takes into account the effect of the neutrino oscillation (See Eq.~(\ref{eq:oscillationfactor})).
We take normal hierarchy in left figure and inverted hierarchy in right figure.
In both figure, we take $\d=0$.
Red, green and blue lines show the spectrum of $\n_e + \bar \n_e$, $\n_\m + \bar \n_\m$ and $\n_\t + \bar \n_\t$, respectively.
}\label{fig:pythiamixed}
\end{figure}

We have calculated the energy spectrum of neutrinos $dN_\n / dE_\n$ from
decay of $N_1$ by using PYTHIA 8.1~\cite{Sjostrand:2007gs}.  The
neutrino spectrum for $M_1 = 2.3~{\rm PeV}$ is shown in
Fig.~\ref{fig:pythia}.  We have a sharp peak in the neutrino energy
spectrum at $E_\n = M_1/2$.  In the case of inverted hierarchy, since
the fractions of muon and tau are large compared to the normal
hierarchy, the number of neutrinos is slightly larger around $E_\nu \sim
10^{5-6}$ GeV due to the decay products of the muons and taus.
%
%

As the neutrino travels towards the Earth, the neutrinos change their
flavors by the neutrino oscillation according to the following
probabilities:
\begin{align}
P(\n_\ell \to \n_{\ell'}) = 
P(\bar\n_\ell \to \bar\n_{\ell'}) &= \sum_{i=1}^3 |U_{\ell i}U_{\ell' i} |^2 \nonumber\\
 &\simeq
\left(\begin{array}{ccc}
0.55 & 0.27 + 0.02\cos\d & 0.18 - 0.02\cos\d \\
0.27 + 0.02\cos\d & 0.36 - 0.02\cos\d & 0.37 + 0.00 \cos\d\\
0.18 - 0.02\cos\d & 0.37 + 0.00\cos\d & 0.45 + 0.02 \cos\d
\end{array}\right). \label{eq:oscillationfactor}
\end{align}
In Fig.~\ref{fig:pythiamixed}, we show the energy spectrum of the
neutrinos after the oscillation.
%

We estimate the observed flux of neutrinos on the Earth in the following
way.
We have two classes of contribution from the decaying dark matter; one
is from halo of our galaxy, and another is from extra-galactic.  For a
review of the calculation of the neutrino flux, {\it e.g.}, see
Ref.~\cite{Grefe:2008zz}.  The halo contribution which is averaged over
the full sky is proportional to $dN_\n /dE_\n$:
\begin{align}
\frac{d\Phi_{\rm halo}}{dE_\n} &= D_{\rm halo} \frac{dN_\n}{dE_\n},
\end{align}
where $D_{\rm halo}$ is determined by the halo density profile $\rho_{\rm halo}(r)$,
\begin{align}
D_{\rm halo} &= \frac{1}{4\pi} \int_{-1}^1 d\sin \theta \int_0^{2\pi} d\phi \left(  \frac{1}{4\pi M_1 \t_{\rm N_1}} \int_0^\infty ds ~\rho_{\rm halo}(r(s,\theta,\phi)) \right). \label{eq:neutrino_from_halo}
\end{align}
The parameter $s$ in the integral of Eq.~(\ref{eq:neutrino_from_halo}) is the distance from the Earth,
and it is related to the distance $r$ from the galactic center as,
$r(s,\theta,\phi) = \sqrt{s^2 + R_\odot^2 - 2sR_\odot \cos\theta \cos\phi}$.
Here, $R_\odot$ is the distance of the Sun to the galactic center, and we take its value as $8.0~{\rm kpc}$ \cite{Gillessen:2008qv}.
For the calculation of $D_{\rm halo}$, we adopt the Navarro-Frenk-White (NFW) density profile \cite{Navarro:1996gj},
\begin{align}
\rho_{\rm halo}(r) &= \rho_\odot \frac{(R_\odot/r_c) (1+R_\odot/r_c)^2 }{(r/r_c) (1+r/r_c)^2},
\end{align}
and take $r_c = 20~{\rm kpc}$ and $\rho_\odot = 0.4~{\rm GeV}~{\rm cm}^{-3}$ \cite{Iocco:2011jz}.
Then, $D_{\rm halo}$ is calculated as,
\begin{align}
D_{\rm halo} &= 1.7 \times 10^{-13} \left(\frac{1~{\rm PeV}}{M_1} \right) \left( \frac{10^{28} {\rm s}}{\t_{\rm N_1}} \right)~{\rm cm}^{-2}{\rm s}^{-1}{\rm sr}^{-1}.
\end{align}
Extra galactic contribution is redshifted because of the expansion of the Universe.
Their contribution is written by,
\begin{align}
\frac{d\Phi_{\rm eg}}{dE_\n} &= \frac{\Omega_{\rm DM} \rho_c c}{4\pi M_1 \t_{N_1}} \int_0^{\infty}
\frac{dz}{H(z)} e^{-s(E_\n,z)} \frac{dN_\n}{dE}\biggr|_{E=(1+z)E_\n}, \label{eq:neutrino_from_eg}
\end{align}
where we estimate the integrand just from $z=0$ to $z_{\rm eq}$ for simplicity and hence
also neglect the contribution from the dark matter which had decayed at the radiation dominated era,
because we assume the dark matter mass is around PeV and the energy of neutrino from early universe is too low to explain the IceCube excess.
In Eq.~(\ref{eq:neutrino_from_eg}), $H(z) = H_0 \sqrt{\Omega_\L + \Omega_m (1+z)^3}$ is the Hubble expansion rate at the redshift $z$.
$c = 3.0 \times 10^{10}~{\rm cm}~{\rm s}^{-1}$ is the speed of light.
$s(E_\n,z)$ is neutrino opacity, which is estimated as $s(E_\n,z) \sim 10^{-17} (1+z)^{7/2} (E_\n / 1{\rm TeV})$ for $z < z_{\rm eq}$ \cite{Esmaili:2012us}.
However, in the present situation, this effect is negligibly small. Then, we take $s(E_\n, z)$ to be zero for an approximation.
For the cosmological parameters, we take $\Omega_\L = 0.68$, $\Omega_m = 0.32$, $\Omega_{\rm DM} = 0.27$, $H_0 = 67$ km s$^{-1}$ Mpc$^{-1}$,
$\rho_c = 3H_0^2 M_{\rm Pl}^2 \simeq 4.7 \times 10^{-6}~{\rm GeV}~{\rm cm}^{-3}$ and $z_{\rm eq} = 3.4 \times 10^3$.
These values are derived from the Planck data \cite{Ade:2013zuv}.

\begin{figure}[t]
\centering
\includegraphics{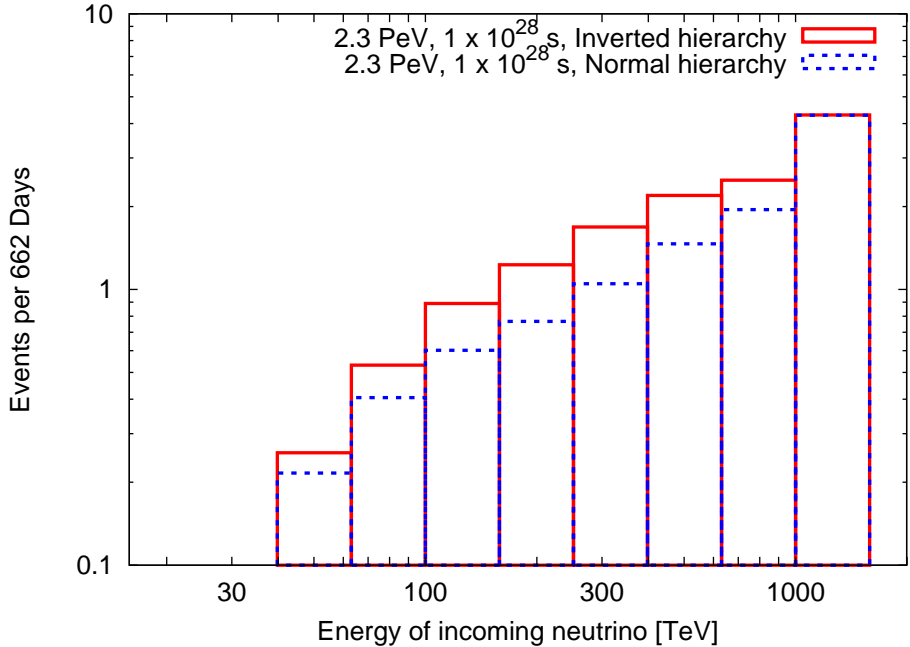}
\caption{
Number of events per 662 days at the IceCube experiment from neutrinos with given energy.
For the parameter of $N_1$, we take $M_1=2.3~{\rm PeV}$, $\t_{N_1} = 10^{28}$ s and CP-violating phase $\d=0$.
We assume normal hierarchy for dotted blue boxes and inverted hierarchy for solid red boxes.
}\label{fig:icecube}
\end{figure}

Finally, the expected number of events at the IceCube detector per 662
days with given energy is calculated as,
\begin{align}
N(E_0 \leq E \leq E_1) = 4\pi \times 662~{\rm days} \times
 \sum_{\ell=e,\m,\t} \int_{E_0}^{E_1} dE_\n \left( \frac{d\Phi_{\rm halo}^{(\nu_\ell + \bar\nu_\ell)}}{dE_\n} + \frac{d\Phi_{\rm eg}^{(\nu_\ell + \bar\nu_\ell)}}{dE_\n} \right) \s_{\rm eff}^{(\n_\ell)}(E_\n),
\end{align}
where $\s_{\rm eff}^{(\nu_\ell)}$ is the neutrino effective area for
each flavor which is given in Refs. \cite{Aartsen:2013jdh,
icecube_data}.  The IceCube experiment observed 28 events with deposited
energies between 30 and 1200 TeV, and the expected number of events from
atmospheric muons and neutrinos is $10.6^{+5.0}_{-3.6}$
\cite{Aartsen:2013jdh}.  For 2.3 PeV dark matter, the total expected
number of events for each pattern of the neutrino mass hierarchy is,
\begin{align}
N(30~{\rm TeV} \leq E_\n) &= 10.8 \times \left( \t_{N_1}/10^{28}~{\rm s} \right)^{-1},\qquad({\rm Normal})\\
N(30~{\rm TeV} \leq E_\n) &= 13.7 \times \left( \t_{N_1}/10^{28}~{\rm s} \right)^{-1}.\qquad({\rm Inverted})
\end{align}
From this estimate, we see that the total excess can be explained for
$\t_{N_1} \simeq 1\times 10^{28}$ s for both normal and inverted hierarchy.
We also show the energy distribution of the neutrinos in Fig.~\ref{fig:icecube}.  In this
figure, we take $M_1= 2.3$~PeV and $\t_{N_1} = 10^{28}$~s.

The IceCube experiment provides the data of the event rate per the
deposited energies in the detector in Fig.~4 in
Ref.~\cite{Aartsen:2013jdh}.
Note that our results in Fig.~\ref{fig:icecube} are, in contrast, those
for incoming neutrino energies, and thus the deposited ones should be
smaller due to escaping neutrinos and muons. One needs to take into
account the correction when the shape of the distribution is compared.
%
%
For $M_{N_1} = 2.3$~PeV, the expected number of neutrinos with the
energy higher than 1~PeV is,
\begin{align}
N(1000~{\rm TeV} \leq E_\n) &= 4.3 \times \left( \t_{N_1}/10^{28}~{\rm s} \right)^{-1},
\end{align}
for both normal and inverted hierarchy.  Thus, by assuming that the
deposited energy is equal to that of incoming neutrinos, the two
observed neutrino events around PeV energies can be explained for
$\t_{N_1} \simeq 2 \times 10^{28}$ s.  We can expect more sub-PeV events
for the inverted hierarchy than the normal hierarchy.  Implications from
the IceCube experiment will be important to distinguish neutrino models.

\section{Summary}

In this paper, we considered a minimalistic cosmological scenario based
on the $U(1)_{B-L}$ extended Standard Model. The model consistently
explains the neutrino masses, the inflation, the baryon asymmetry of the
Universe and dark matter abundance, which are left unexplained in the
Standard Model.  
If both the baryon asymmetry and the dark matter abundance are
explained directly by the inflaton decay,
we obtain the mass of $N_1$ and the second lightest right-handed neutrino $N_2$ to be 1 PeV and $10^{12}$ GeV, respectively.
%

Interestingly, the mass of $N_1$, PeV, turns out to be the energy scale
of the excess of the neutrino events at the IceCube experiment.
We see that the PeV neutrino events can be
explained by the decaying $N_1$ with its lifetime being ${\cal
O}(10^{28})$~s.  Predicted number of neutrinos with sub-PeV energies
depends on the neutrino mass hierarchy and the CP violating
phase. Further observations of high energy neutrino events may, in
principle, provide information on the flavor structures in the neutrino
sector.
%

If the coupling between the Standard Model Higgs and the $B-L$ Higgs
field is significant, the reheating temperature can be higher than the
second lightest right-handed neutrino $N_2$, depending on the
coupling. In such a case, thermal leptogenesis is possible whereas the dark matter
should be heavier than ${\cal O}(10)$ PeV to explain the abundance by the inflaton decay.

\section*{Acknowledgement}
We would like to thank Aya Ishihara for the explanation of the IceCube
experiments. This work is supported by JSPS Grant-in-Aid for Young
Scientists (B) (No.~23740165 [RK], No.~25800169 [TH]), MEXT Grant-in-Aid
for Scientific Research on Innovative Areas (No.~25105011 [RK]) and JSPS
Research Fellowships for Young Scientists [RS].


\begin{thebibliography}{1}


\bibitem{Guth:1980zm}
  A.~H.~Guth,
  Phys.\ Rev.\  D {\bf 23}, 347-356 (1981).
\bibitem{Starobinsky:1980te}
A.~A.~Starobinsky,
Phys.\ Lett.\ B {\bf 91} (1980) 99.
%
\bibitem{Sato:1980yn}
  K.~Sato,
  Mon.\ Not.\ Roy.\ Astron.\ Soc.\  {\bf 195}, 467-479 (1981).




\bibitem{Okada:2013vxa} 
  N.~Okada and Q.~Shafi,
arXiv:1311.0921 [hep-ph].  

\bibitem{Okada:2014lxa} 
  N.~Okada, V.~N.~Senoguz and Q.~Shafi,
arXiv:1403.6403 [hep-ph].  

\bibitem{Linde:1983gd}
  A.~D.~Linde,
  Phys.\ Lett.\ B {\bf 129}, 177 (1983).


\bibitem{Minkowski:1977sc} 
  P.~Minkowski,
  Phys.\ Lett.\ B {\bf 67}, 421 (1977);
%
T.~Yanagida, in Proceedings of the Workshop on Unified Theory and Baryon Number of the Universe, eds. O.~Sawada and A.~Sugamoto (KEK, 1979) p.95;
%
M.~Gell-Mann, P.~Ramond and R.~Slansky, in Supergravity, ed. by D.~Freedman and P.~Van Nieuwenhuizen, North Holland, Amsterdam (1979), pp.~315-321;
%
S.~Glashow, in Quarks and Leptons, Carg$\grave{\rm e}$se 1979, eds.~M.~L$\acute{\rm e}$vy et al., (Plenum, 1980, New York);
%
  R.~N.~Mohapatra and G.~Senjanovic,
  Phys.\ Rev.\ Lett.\  {\bf 44}, 912 (1980);
  J.~Schechter and J.~W.~F.~Valle,
  Phys.\ Rev.\ D {\bf 22}, 2227 (1980).


\bibitem{Fukugita:1986hr} 
  M.~Fukugita and T.~Yanagida,
  Phys.\ Lett.\ B {\bf 174}, 45 (1986).

\bibitem{Davoudiasl:2004be} 
  H.~Davoudiasl, R.~Kitano, T.~Li and H.~Murayama,
  Phys.\ Lett.\ B {\bf 609}, 117 (2005)
  [hep-ph/0405097].


\bibitem{Shafi:1983bd} 
  Q.~Shafi and A.~Vilenkin,
Phys.\ Rev.\ Lett.\  {\bf 52}, 691 (1984);
 %
  Q.~Shafi and V.~N.~Senoguz,
Phys.\ Rev.\ D {\bf 73}, 127301 (2006)  [astro-ph/0603830];
%
  C.~Destri, H.~J.~de Vega and N.~G.~Sanchez,
Phys.\ Rev.\ D {\bf 77}, 043509 (2008)  [astro-ph/0703417];
%
  R.~Kallosh and A.~D.~Linde,
JCAP {\bf 0704}, 017 (2007)  [arXiv:0704.0647 [hep-th]];
%
  T.~L.~Smith, M.~Kamionkowski and A.~Cooray,
Phys.\ Rev.\ D {\bf 78}, 083525 (2008)  [arXiv:0802.1530 [astro-ph]];
%
  V.~N.~Senoguz and Q.~Shafi,
Phys.\ Lett.\ B {\bf 668}, 6 (2008)  [arXiv:0806.2798 [hep-ph]]; 
%
  M.~U.~Rehman, Q.~Shafi and J.~R.~Wickman,
 Phys.\ Rev.\ D {\bf 78}, 123516 (2008)  [arXiv:0810.3625 [hep-ph]]; 
%
  M.~U.~Rehman and Q.~Shafi,
Phys.\ Rev.\ D {\bf 81}, 123525 (2010)  [arXiv:1003.5915 [astro-ph.CO]].  


\bibitem{Nakayama:2010sk} 
  K.~Nakayama and F.~Takahashi,
JCAP {\bf 1102}, 010 (2011)  [arXiv:1008.4457 [hep-ph]];
%
  K.~Nakayama and F.~Takahashi,
arXiv:1403.4132 [hep-ph];
%
see also
  K.~Nakayama and F.~Takahashi,
JCAP {\bf 1011}, 009 (2010)  [arXiv:1008.2956 [hep-ph]]. 

\bibitem{Hamada:2014iga} 
  Y.~Hamada, H.~Kawai, K.~-y.~Oda and S.~C.~Park,
arXiv:1403.5043 [hep-ph];
%
  F.~Bezrukov and M.~Shaposhnikov,
arXiv:1403.6078 [hep-ph];
%
see also
  J.~L.~Cook, L.~M.~Krauss, A.~J.~Long and S.~Sabharwal,
arXiv:1403.4971 [astro-ph.CO];
%
  F.~L.~Bezrukov and M.~Shaposhnikov,
Phys.\ Lett.\ B {\bf 659}, 703 (2008)  [arXiv:0710.3755 [hep-th]].  


\bibitem{Peebles:1982ib} 
  P.~J.~E.~Peebles,
 Astrophys.\ J.\  {\bf 258}, 415 (1982); 
%
  K.~A.~Olive and M.~S.~Turner,
Phys.\ Rev.\ D {\bf 25}, 213 (1982).  


\bibitem{Dodelson:1993je} 
  S.~Dodelson and L.~M.~Widrow,
Phys.\ Rev.\ Lett.\  {\bf 72}, 17 (1994)  [hep-ph/9303287].  
%
  X.~-D.~Shi and G.~M.~Fuller,
Phys.\ Rev.\ Lett.\  {\bf 82}, 2832 (1999)  [astro-ph/9810076];
%
  A.~D.~Dolgov and S.~H.~Hansen,
 Astropart.\ Phys.\  {\bf 16}, 339 (2002)  [hep-ph/0009083];
%
  K.~Abazajian, G.~M.~Fuller and M.~Patel,
Phys.\ Rev.\ D {\bf 64}, 023501 (2001)  [astro-ph/0101524]; 

\bibitem{Asaka:2005an} 
  T.~Asaka, S.~Blanchet and M.~Shaposhnikov,
Phys.\ Lett.\ B {\bf 631}, 151 (2005)  [hep-ph/0503065];
%
  T.~Asaka, M.~Laine and M.~Shaposhnikov,
 JHEP {\bf 0701}, 091 (2007)  [hep-ph/0612182]. 


\bibitem{Kusenko:2010ik} 
  A.~Kusenko, F.~Takahashi and T.~T.~Yanagida,
Phys.\ Lett.\ B {\bf 693}, 144 (2010)  [arXiv:1006.1731 [hep-ph]];  
%
  H.~Ishida, K.~S.~Jeong and F.~Takahashi,
Phys.\ Lett.\ B {\bf 731}, 242 (2014)  [arXiv:1309.3069 [hep-ph]].  



\bibitem{Aartsen:2013bka} 
  M.~G.~Aartsen {\it et al.}  [IceCube Collaboration],
  Phys.\ Rev.\ Lett.\  {\bf 111}, no. 2, 021103 (2013)
  [arXiv:1304.5356 [astro-ph.HE]].

\bibitem{Aartsen:2013jdh} 
  M.~G.~Aartsen {\it et al.}  [IceCube Collaboration],
  Science {\bf 342}, no. 6161, 1242856 (2013)
  [arXiv:1311.5238 [astro-ph.HE]].


\bibitem{Feldstein:2013kka} 
  B.~Feldstein, A.~Kusenko, S.~Matsumoto and T.~T.~Yanagida,
  Phys.\ Rev.\ D {\bf 88}, no. 1, 015004 (2013)
  [arXiv:1303.7320 [hep-ph]].

\bibitem{Esmaili:2013gha} 
  A.~Esmaili and P.~D.~Serpico,
  JCAP {\bf 1311}, 054 (2013)
  [arXiv:1308.1105 [hep-ph]].


\bibitem{Banks:2010zn}
  T.~Banks and N.~Seiberg,
Phys.\ Rev.\ D {\bf 83}, 084019 (2011)  [arXiv:1011.5120 [hep-th]].

\bibitem{BerasaluceGonzalez:2011wy}
  M.~Berasaluce-Gonzalez, L.~E.~Ibanez, P.~Soler and A.~M.~Uranga,
JHEP {\bf 1112}, 113 (2011)  [arXiv:1106.4169 [hep-th]];
%
  M.~Berasaluce-Gonzalez, P.~G.~Camara, F.~Marchesano, D.~Regalado and A.~M.~Uranga,
JHEP {\bf 1209}, 059 (2012)  [arXiv:1206.2383 [hep-th]];
%
see also  R.~Blumenhagen, M.~Cvetic, S.~Kachru and T.~Weigand,
Ann.\ Rev.\ Nucl.\ Part.\ Sci.\  {\bf 59}, 269 (2009)
[arXiv:0902.3251 [hep-th]].



\bibitem{'tHooft:1976fv} 
  G.~'t Hooft,
 Phys.\ Rev.\ D {\bf 14}, 3432 (1976)  [Erratum-ibid.\ D {\bf 18}, 2199 (1978)]; 
%
  G.~'t Hooft,
Phys.\ Rept.\  {\bf 142}, 357 (1986). 
 

\bibitem{Frampton:2002qc} 
  P.~H.~Frampton, S.~L.~Glashow and T.~Yanagida,
  Phys.\ Lett.\ B {\bf 548}, 119 (2002)
  [hep-ph/0208157].

\bibitem{Harigaya:2012bw} 
  K.~Harigaya, M.~Ibe and T.~T.~Yanagida,
  Phys.\ Rev.\ D {\bf 86}, 013002 (2012)
  [arXiv:1205.2198 [hep-ph]].


\bibitem{Pontecorvo:1957qd} 
  B.~Pontecorvo,
  Sov.\ Phys.\ JETP {\bf 7}, 172 (1958)
  [Zh.\ Eksp.\ Teor.\ Fiz.\  {\bf 34}, 247 (1957)].
\bibitem{MNS}
  Z.~Maki, M.~Nakagawa and S.~Sakata,
  Prog.\ Theor.\ Phys.\  {\bf 28}, 870 (1962).



\bibitem{Casas:2001sr} 
  J.~A.~Casas and A.~Ibarra,
  Nucl.\ Phys.\ B {\bf 618}, 171 (2001)
  [hep-ph/0103065].

\bibitem{Ibarra:2003up} 
  A.~Ibarra and G.~G.~Ross,
  Phys.\ Lett.\ B {\bf 591}, 285 (2004)
  [hep-ph/0312138].


\bibitem{Liddle:2000cg} 
  A.~R.~Liddle and D.~H.~Lyth,
Cambridge, UK: Univ. Pr. (2000) 400 p  


\bibitem{Ade:2013uln} 
  P.~A.~R.~Ade {\it et al.}  [Planck Collaboration],
 arXiv:1303.5082 [astro-ph.CO].  


\bibitem{Ade:2014xna} 
  P.~A.~R.~Ade {\it et al.}  [BICEP2 Collaboration],
  arXiv:1403.3985 [astro-ph.CO].


\bibitem{Giusarma:2014zza} 
  E.~Giusarma, E.~Di Valentino, M.~Lattanzi, A.~Melchiorri and O.~Mena,
  arXiv:1403.4852 [astro-ph.CO].

\bibitem{Contaldi:2014zua} 
  C.~R.~Contaldi, M.~Peloso and L.~Sorbo,
arXiv:1403.4596 [astro-ph.CO].  

\bibitem{Kawasaki:2014lqa} 
  M.~Kawasaki and S.~Yokoyama,
  arXiv:1403.5823 [astro-ph.CO];
%
  M.~Kawasaki, T.~Sekiguchi, T.~Takahashi and S.~Yokoyama,
arXiv:1404.2175 [astro-ph.CO]. 

  
\bibitem{Miranda:2014wga} 
  V.~Miranda, W.~Hu and P.~Adshead,
  arXiv:1403.5231 [astro-ph.CO];
%
see also  B.~Feng and X.~Zhang,
  Phys.\ Lett.\ B {\bf 570}, 145 (2003)
  [astro-ph/0305020];
%
  M.~Kawasaki and F.~Takahashi,
  Phys.\ Lett.\ B {\bf 570}, 151 (2003)
  [hep-ph/0305319].

\bibitem{Freivogel:2014hca} 
  B.~Freivogel, M.~Kleban, M.~R.~Martinez and L.~Susskind,
arXiv:1404.2274 [astro-ph.CO];
%
  R.~Bousso, D.~Harlow and L.~Senatore,
 arXiv:1404.2278 [astro-ph.CO];
%
see also  H.~Murayama, K.~Nakayama, F.~Takahashi and T.~T.~Yanagida,
arXiv:1404.3857 [hep-ph];
%
  T.~Higaki and F.~Takahashi,
 arXiv:1404.6923 [hep-th].  



\bibitem{Mukaida:2012qn} 
  K.~Mukaida and K.~Nakayama,
JCAP {\bf 1301}, 017 (2013)  
[arXiv:1208.3399 [hep-ph]]; 
%
  K.~Mukaida and K.~Nakayama,
JCAP {\bf 1303}, 002 (2013)  [arXiv:1212.4985 [hep-ph]].  


\bibitem{Asaka:1999yd} 
  T.~Asaka, K.~Hamaguchi, M.~Kawasaki and T.~Yanagida,
Phys.\ Lett.\ B {\bf 464}, 12 (1999)  [hep-ph/9906366];  
%
  T.~Asaka, K.~Hamaguchi, M.~Kawasaki and T.~Yanagida,
Phys.\ Rev.\ D {\bf 61}, 083512 (2000)  [hep-ph/9907559].  



\bibitem{Covi:1996wh} 
  L.~Covi, E.~Roulet and F.~Vissani,
  Phys.\ Lett.\ B {\bf 384}, 169 (1996)
  [hep-ph/9605319].



\bibitem{Davidson:2002qv} 
  S.~Davidson and A.~Ibarra,
  Phys.\ Lett.\ B {\bf 535}, 25 (2002)
  [hep-ph/0202239].



\bibitem{Beringer:1900zz} 
  J.~Beringer {\it et al.}  [Particle Data Group Collaboration],
  Phys.\ Rev.\ D {\bf 86}, 010001 (2012).


\bibitem{Ade:2013zuv} 
  P.~A.~R.~Ade {\it et al.}  [Planck Collaboration],
arXiv:1303.5076 [astro-ph.CO].  


\bibitem{Buchmuller:2002rq} 
  W.~Buchmuller, P.~Di Bari and M.~Plumacher,
Nucl.\ Phys.\ B {\bf 643}, 367 (2002)  [Erratum-ibid.\ B {\bf 793}, 362 (2008)]  [hep-ph/0205349]. 



\bibitem{Cornwall:1974km}
  J.~M.~Cornwall, D.~N.~Levin and G.~Tiktopoulos,
  Phys.\ Rev.\  D {\bf 10}, 1145 (1974)
  [Erratum-ibid.\  D {\bf 11}, 972 (1975)];
  C.~E.~Vayonakis,
  Lett.\ Nuovo Cim.\  {\bf 17}, 383 (1976);
  B.~W.~Lee, C.~Quigg and H.~B.~Thacker,
  Phys.\ Rev.\ D {\bf 16}, 1519 (1977);
  G.~J.~Gounaris, R.~Kogerler and H.~Neufeld,
  Phys.\ Rev.\ D {\bf 34}, 3257 (1986).


\bibitem{Sjostrand:2007gs} 
  T.~Sjostrand, S.~Mrenna and P.~Z.~Skands,
  Comput.\ Phys.\ Commun.\  {\bf 178}, 852 (2008)
  [arXiv:0710.3820 [hep-ph]].


\bibitem{Grefe:2008zz} 
  M.~Grefe,
  arXiv:1111.6041 [hep-ph].



\bibitem{Gillessen:2008qv} 
  S.~Gillessen, F.~Eisenhauer, S.~Trippe, T.~Alexander, R.~Genzel, F.~Martins and T.~Ott,
  Astrophys.\ J.\  {\bf 692}, 1075 (2009)
  [arXiv:0810.4674 [astro-ph]].


\bibitem{Navarro:1996gj} 
  J.~F.~Navarro, C.~S.~Frenk and S.~D.~M.~White,
  Astrophys.\ J.\  {\bf 490}, 493 (1997)
  [astro-ph/9611107].



\bibitem{Iocco:2011jz} 
  F.~Iocco, M.~Pato, G.~Bertone and P.~Jetzer,
  JCAP {\bf 1111}, 029 (2011)
  [arXiv:1107.5810 [astro-ph.GA]].


\bibitem{Esmaili:2012us} 
  A.~Esmaili, A.~Ibarra and O.~L.~G.~Peres,
  JCAP {\bf 1211}, 034 (2012)
  [arXiv:1205.5281 [hep-ph]].



\bibitem{icecube_data}
http://www.icecube.wisc.edu/science/data

\end{thebibliography}
\end{document}